\documentclass[aps,a4paper,superscriptaddress,preprint,showpacs,showkeys,12pt]{revtex4-1}
\usepackage{graphicx}
\usepackage{subfigure}
\usepackage{latexsym}   

\begin{document}

\title{Multicanonical entropy like-solution of statistical temperature weighted histogram
analysis method (ST-WHAM)}
\author{\firstname{Leandro} G. \surname{Rizzi}}
\email{lerizzi@usp.br} 
\author{\firstname{Nelson} A. \surname{Alves}}
\email{alves@ffclrp.usp.br}
\affiliation{Departamento de F\'{\i}sica, FFCLRP, Universidade de S\~ao Paulo,
             Avenida Bandeirantes, 3900, 14040-901, Ribeir\~ao Preto, SP, Brazil.}
\date{\today}


\begin{abstract}
  	A multicanonical update relation for calculation of the microcanonical
entropy $S_{micro}(E)$ by means of the estimates of the inverse statistical temperature $\beta_S$,
is proposed. 
   This inverse temperature is obtained from the recently proposed	
statistical temperature weighted histogram analysis method (ST-WHAM).	
		The performance of ST-WHAM concerning the computation of $S_{micro}(E)$ from canonical measures, 
in a model with strong free-energy barriers, is also discussed on the basis of comparison with the multicanonical simulation estimates.
\end{abstract}

\keywords{WHAM equations, ST-WHAM equation, Dipolar Ising model, Multicanonical simulation}
\pacs{05.10.-a, 05.10.Ln}

\maketitle


  Reweighting and histogram methods 
	\cite{salsburg,singer,valleau,bennett,falcioni,bhanot,ferrenberg-rewei,wham,alves90,alves92}
have become essential tools for calculation of thermodynamic averages from 
samplings obtained with different coupling parameters. 
  The multiple histogram method \cite{wham,bereau}, also known as weighted histogram analysis method (WHAM),
has greatly improved the efficiency of Monte Carlo simulations \cite{kumar1,kumar2},
mainly when it comes to dealing with data obtained from replica-exchange method (REM) at different temperatures 
{\cite{okamoto1,okamoto2}.
  The REM simulation has been particularly important to sample conformations in protein folding 
simulations.
  Generalized ensemble algorithms, in which REM, the multicanonical algorithm (MUCA) and their 
extensions \cite{okamoto3,okamoto4,straub130,straub132,straub133} are included,
allow for the simulation to overcome the free-energy barriers frequently encountered in the energy landscapes.
	
  The WHAM equations combine data from an arbitrary number $M$ of independent Monte Carlo or 
molecular dynamics simulations to enhance the sampling, thereby producing
thermal averages as a continuous function of the coupling parameter, being the temperature 
the most often parameter.
  The solutions of the WHAM equations, obtained from energy data stored in histograms 
$H_{\alpha}(E)$, $\alpha = 1, \cdots, M$ yield free-energy differences via an iterative 
numerical process.
  The success of the iterative process depends on the number of different histograms and their
overlaps \cite{bereau}.
  More recently \cite{st-wham}, an iteration-free approach to solve the WHAM equations
in terms of intensive variables has been developed. 
  This numerical approach, namely the statistical temperature weighted histogram analysis method (ST-WHAM),
yields the inverse temperature $\beta_S= \partial S/ \partial E$ directly from a new 
form of WHAM equations.
   Thermodynamic quantities like entropy, can be evaluated upon numerical integration of this 
statistical temperature.

   The formulation of the new WHAM equations now starts from a weighted average of the
numerical estimates for the density of states $\Omega_{\alpha}$, 
\begin{equation}
   \tilde{\Omega}(E) = \sum_{\alpha}^M \tilde{f}_{\alpha}(E)  \Omega_{\alpha}(E)  ,  
                                                                           \label{eq:st-wham1}
\end{equation}
with the normalization condition $\sum_{\alpha}\tilde{f}_{\alpha}(E) =1 $.
	  The density of states are estimated from the energy histograms $H_{\alpha}$,
$\Omega_{\alpha}(E) = H_{\alpha}/\Pi_{\alpha}$, where
$\Pi_{\alpha} = N_{\alpha} W_{\alpha}(E)/Z_{\alpha}$, as consequence of the WHAM formulation
\cite{st-wham}.
    The quantities $N_{\alpha}, W_{\alpha}$, and $ Z_{\alpha}$ are the number of energy entries in the
histograms $H_{\alpha}$, the sampling weights, and the (unknown) partition functions, respectively.
    The final weighted temperature is given by
\begin{eqnarray}
 \tilde{\beta}_S(E) & = &\frac{\partial\, {\rm ln} \tilde{\Omega}(E)}{\partial E} \label{eq:st-wham2}\\  
               & = & \sum_{\alpha} f_{\alpha}^*(\beta_{\alpha}^H + \beta_{\alpha}^W) + 
				\sum_{\alpha} f_{\alpha}^* 
					\frac{\partial\, {\rm ln} \tilde{f}_{\alpha}}{\partial E}  ,
		\label{eq:st-wham3}
\end{eqnarray}
where $f_{\alpha}^*     =H_{\alpha}/\sum_{\alpha}H_{\alpha}$, 
      $\beta_{\alpha}^H = \partial\, {\rm ln} H_{\alpha}/{\partial E}$,  
      $\beta_{\alpha}^W= -\partial\, {\rm ln} W_{\alpha}/{\partial E}$, and
			$\tilde{f}_{\alpha} = \Pi_{\alpha}/\sum_{\alpha} \Pi_{\alpha}$.
 
  The ST-WHAM approach estimates the thermodynamic temperature from the first term
in Eq. (\ref{eq:st-wham3}) only. This is because the second term, which amounts to the difference between
ST-WHAM and WHAM, can be neglected for large samplings, $ N_{\alpha} >> 1$.
   This remark is what makes the ST-WHAM an iteration-free method to estimate the inverse effective temperature $\tilde{\beta}_S$.
    Now, entropy estimates follow from a careful integration of  $\tilde{\beta}_S(E)$
when one disregards the second term in Eq. (\ref{eq:st-wham3}),
\begin{equation}
   \tilde{S}(E) = \int{  \sum_{\alpha} f_{\alpha}^*(\beta_{\alpha}^H + \beta_{\alpha}^W)\, dE} .
	                                                                                  \label{eq:ST}
\end{equation}

{\bf Microcanonical analysis and aims - }
   Two sampling algorithms are successfully used to estimate the density of states
$\Omega(E)$: the  Wang-Landau algorithm \cite{WL1,WL2} and MUCA \cite{Berg2001,Berg2003}.
	 This success is intimately related to their performances in obtaining a reasonable number
of round-trips between two extremal energy values \cite{round-trip}.
   Thus, depending on whether the system presents strong free-energy barrier, or not,
we may face some failure.
   Of course this is not a feature of these algorithms only, but in fact it is an overall
behavior even for other generalized-ensemble algorithms \cite{straub132}.
   Algorithms like MUCA facilitate the microcanonical analysis, which is important
for characterization of the thermodynamic aspects of phase transition in small systems.
   Among the MUCA applications and microcanonical analysis are the studies of 
heteropolymer aggregations \cite{janke-L2006,janke-C2008,janke-C2009,frigori-2011}.

    The microcanonical analysis contrasts with the usual data analysis obtained from
simulations in the canonical ensemble.
    For example, it is possible to observe thermodynamic features like temperature 
discontinuity and negative specific heat in the microcanonical ensemble  
\cite{Ruffo_PRL2001,TurkingtoJSP101,kastner_A2007,Ruffo_EPJB2008,Ruffo_Rep2009,Ellis_A335},
which appear at first-order phase transitions.
		A negative specific heat is a consequence of the nonconcave behavior of the 
microcanonical entropy $S_{micro}(E)$ as a function of the energy and leads to the 
so-called convex intruder in $S_{micro}(E)$ \cite{gross-book}, a feature 
that is prohibited in the canonical ensemble. 
    Systems where one finds convex intruders in the entropy present ensemble inequivalence 
\cite{TurkingtoJSP101,ruffo-2002,costeniuc-2006}.
    It is noteworthy that the functions entropy $S(E)$ and free-energy
$F(\beta)$ are related by the Legendre-Fenchel (LF) transform, $F(\beta) = {\cal L}[S(E)]$.
		This relation is always true, independent of the shape of $S(E)$, even with a nonconcave 
piece in its domain.
    The LF transform always produces a concave function.
    On the other hand, if a function is not concave in its domain, then it cannot
be obtained as the LF transform of another function \cite{costeniuc-2006}.
    Thus, if $S_{micro}(E)$ has a convex intruder, it cannot be calculated from free energies 
in that energy domain and, as a consequence one has thermodynamically nonequivalent systems. 
	  With respect to WHAM, it is known that this method in essence provides
free energies, and this has motivated our comparative study of the entropic results produced 
by ST-WHAM and MUCA.
    In this sense, the present study investigates the performance of the ST-WHAM 
in producing the microcanonical entropy from a set of data obtained with canonical measures.
    Another important aim here is to show that the integral in Eq. (\ref{eq:ST})
can indeed be replaced with the simple iteration relation used for updating the multicanonical 
parameters.
		To this end, data obtained from the two-dimensional (2D) Ising model with 
competitive nearest-neighbor ferromagnetic and long-range dipolar interactions are analyzed.
    This model presents strong free-energy barriers for the coupling 
$\delta=2$, and it becomes stronger as the lattice size increases.
   Indeed, this model should be considered a benchmark for testing the performance of 
optimized MC algorithms.

{\bf Methods - }
	 The multicanonical algorithm consists of sampling configurations with weight 
$w_{mu}(E) \simeq 1/\Omega(E)$. 
	 Therefore, the histogram $H(E) \propto \Omega(E) w_{mu}(E)$ is almost flat.
   If the Boltzmann entropy $S(E)= \ln \Omega(E)$ is considered, it is convenient to
write the following parameterization for the entropy $S(E)=b(E)E-a(E)$, 
where $a(E)$ and $b(E)$ are the multicanonical parameters. 
	 Hence,  the multicanonical weight is given by $w_{mu}(E)={\rm exp}[-b(E)E+a(E)]$, 
with the parameter $a(E)$ related to a multicanonical free energy and $b(E)$ related 
to the inverse of the microcanonical temperature.
	
   The multicanonical parameters $a(E)$ and $b(E)$ are estimated from  $N_{r}$ multicanonical simulations.
   Usually, the number $N_r$ is defined {\it a posteriori} when the multicanonical parameters
present some convergence.
	The implementation of the multicanonical method requires the discretization of the energy.
	 To this end, the labeled energies $E_{m}=E_{0} + m \varepsilon$ are defined, 
where each energy is associated with an integer number $m$. 
   All the energies in the interval $[E_{m},E_{m+1}[$ are in the $m$th energy bin of size 
$\varepsilon$ and contribute to the  histogram $H_{mu}(E_{m})$. 
   The constant $E_{0}$ is defined as a reference energy near but below the ground-state energy.
	 The method consists of updating the multicanonical parameters as follows,
\begin{eqnarray}
   a^{n}(E_{m-1})  & = & a^{n}(E_{m}) + [b^{n}(E_{m-1})-b^{n}(E_{m})]E_{m}~~ \label{eq:muca1} \\
    b^{n}(E_{m})   & = & b^{n-1}(E_{m}) +
[ \ln \hat{H}^{n-1}_{mu}(E_{m+1}) 
		                    -\ln \hat{H}^{n-1}_{mu}(E_{m}) ] / \varepsilon ,     \label{eq:muca2}
\end{eqnarray}
where $n=1, \cdots, N_r$ and $\hat{H}^{n}_{mu}(E_{m})=\max[h_{0},H^{n}_{mu}(E_{m})]$, 
with $0<h_{0}<1$.
   It is convenient to compute the above recurrence relations with the initial conditions 
$a^{0}(E_{m})=0$ and small values for $b^{0}(E_{m})$, if the simulation uses a hot-start initialization.
	 As the multicanonical parameters are updated, the method samples configurations 
with lower energies. 

    Here, the numerical integration in Eq. (\ref{eq:ST}}) is replaced with the iterative MUCA procedure
described in Eq. (\ref{eq:muca1}):  $ \tilde{\beta}_S(E) \rightarrow b(E)$. 
    To illustrate this {\it integration} method, in a system displaying strong free-energy barriers, 
the 2D dipolar Ising model defined on a lattice with $N=L^{2}$ sites occupied 
by Ising spins $\sigma_{i}=\pm 1$ is considered,
\begin{equation}
   {\cal H} = -\delta \sum_{<i,j>} \sigma_{i} \sigma_{j} +  
             \sum_{i < j} \frac{\sigma_{i} \sigma_{j}}{r_{ij}^{3}} \, ,    \label{eq:hamiltonian}
\end{equation}
where $\delta = J/g$ is the ratio between the ferromagnetic exchange interaction ($J>0$) and the dipolar 
antiferromagnetic interaction of strength $g>0$, and $r_{ij}$ is the distance between distinct pairs of
lattice spins $i$ and $j$.

  This dipolar Ising model has been employed to describe thermodynamic properties of ultrathin magnetic films
\cite{Debell2000} and presents three phases for the coupling $\delta=2$
\cite{Cannas2006}.	
    More recently, we have studied the phase transitions for this coupling 
using the Metropolis algorithm and the multiple histogram analysis \cite{Rizzi2010}.

{\bf Numerical results - }
   Next, comparative results from ST-WHAM and MUCA simulations for $L=32$ and 48 are presented.
	 The histograms $H_{\alpha}$ were obtained at five temperatures $T_{\alpha}$ and contain
$3.4 \times 10^7$ measurements.
   The study concerning the multicanonical results are based on  $N_r =10^3$ multicanonical updates 
and $5 \times 10^4$ histogram entries in each simulation, for both $L=32$ and 48
lattice sizes, and with $\varepsilon =1$.

   The Fig. 1(a) shows the histograms $H_{\alpha}$ obtained at $T_{\alpha}$ for $L=32$.
   The multicanonical and ST-WHAM results for the inverse temperature are depicted in Fig. 1(b).
   These two figures reveal a noisy behaviour for $E/N \lesssim -1.18$, 
because both the Metropolis histograms and MUCA do not contain appreciable energy measurements in that region. 
    Entropy calculations use the ST-WHAM estimates for $\tilde{\beta}_S(E)$ followed by the multicanonical
update equation $ a(E_{m-1}) =  a(E_{m}) + [\tilde{\beta}_S(E_{m-1})- \tilde{\beta}_S(E_{m})]E_{m}$.
    This evaluation is denoted by ST-WHAM-MUCA here.
    Since the nonconcavity of the microcanonical entropy can hardly be seen on a plot, we exhibit
in Fig. 1(c) the shifted microcanonical entropy 
$\Delta{S}_{micro}(E) =  S_{\rm micro}(E)- (A + B\varepsilon)$.
    The subtraction is performed for visualization of the nonconcavity of the entropy in relation to 
the linear function joining $S_{\rm micro}(E_a)$ to $S_{\rm micro}(E_b)$, where $E_a$ and $E_b$ 
are such that $\Delta{S}_{micro}(E_a) = \Delta{S}_{micro}(E_b) =0$. 
    The MUCA estimates for $S_{micro}(E)$ are also included in Fig. 1(c), impressively,
both estimates, ST-WHAM-MUCA and MUCA, yield the same results
in the energy range where the phase transition takes place.
    This energy range corresponds to the so-called backbending or S-loop in the caloric curve. 
    We also verified that the numerical integration of Eq. (\ref{eq:ST}) via 
an analytical approximation \cite{st-wham} with energy bin size $\varepsilon =1$
gives the same results as the ones achieved by means of the ST-WHAM-MUCA procedure.

    Figure 2 displays our ST-WHAM and MUCA results for $L=48$.
		Actually, this model has two phase transitions for the coupling $\delta=2$,
which are revealed for large lattice sizes only.
    Figure 2(a) shows the Metropolis histograms $H_{\alpha}$ obtained at 5 temperatures.
		The results for the inverse temperature are depicted in Fig. 2(b).
		Backbendings are observed at $E/N \simeq -1.15$ and $E/N \simeq -1.07$, which
signals the presence of phase transitions.
    They are called stripe-nematic and nematic-tetragonal, respectively.
    This figure shows a noisier behavior for the multicanonical inverse temperature 
as compared to the one obtained for $L=32$, because tunneling events 
between the lower energy phase and the higher energy one are suppressed
due to stronger free-energy barriers.
     As a matter of fact, this noisy behavior provides evidence for the needs of larger 
statistics in the MUCA simulation.
    Figures 2(c) and (d) show the shifted entropies at both phase transitions.
		Again, both methods give results in impressive agreement, mainly if we consider that 
the apparent disagreement between the ST-WHAM and MUCA results may be explained by the 
inefficient Metropolis sampling at low temperatures and in the critical region
of the stripe-nematic phase transition.

	   In summary, we have highlighted the performance of ST-WHAM in obtaining the 
microcanonical entropy, even from only a few energy histograms produced with a canonical 
measure.  
		 A simple update procedure, based on the MUCA equations, can also be readily used
to evaluate $S_{micro}(E)$ from the inverse temperature estimates by employing ST-WHAM, 
which avoids possible numerical instabilities at a first-order phase transition. 
     Therefore, this approach can be used to obtain the whole microcanonical entropy in 
an efficient way, and can stands out from other methods that also evaluate $S_{micro}(E)$ but
require the proper computation of transition matrices \cite{wang-swendsen}.

{\bf Acknowledgments -}
The authors acknowledge support by the Brazilian agencies FAPESP, CAPES, and CNPq.

\begin{figure}[t]
\begin{center}
\includegraphics[height=14cm]{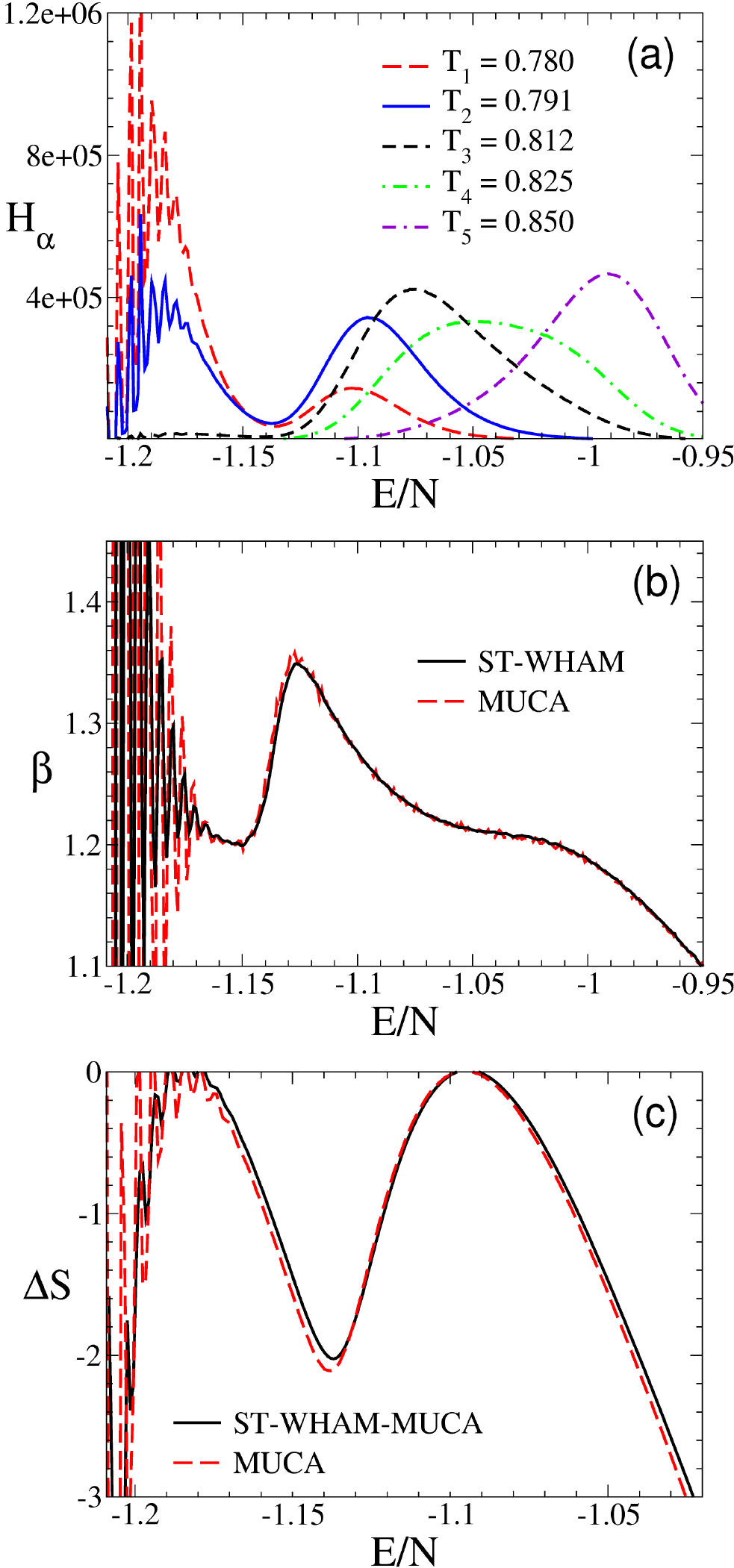}%
\end{center}
\vspace{-0.3cm}
\caption{(a) histograms $H_{\alpha}$; (b) ST-WHAM (solid line) and MUCA (dashed line)
estimates for the inverse temperature; (c) shifted microcanonical entropies estimated from
 multicanonical-like solution of Eq. (\ref{eq:ST}) (solid line) and MUCA
(dashed line) as a function of $E/N$ for lattice size $L=32$.}
\end{figure}

\newpage

\begin{figure}[t]
\begin{center}
\includegraphics[height=14cm]{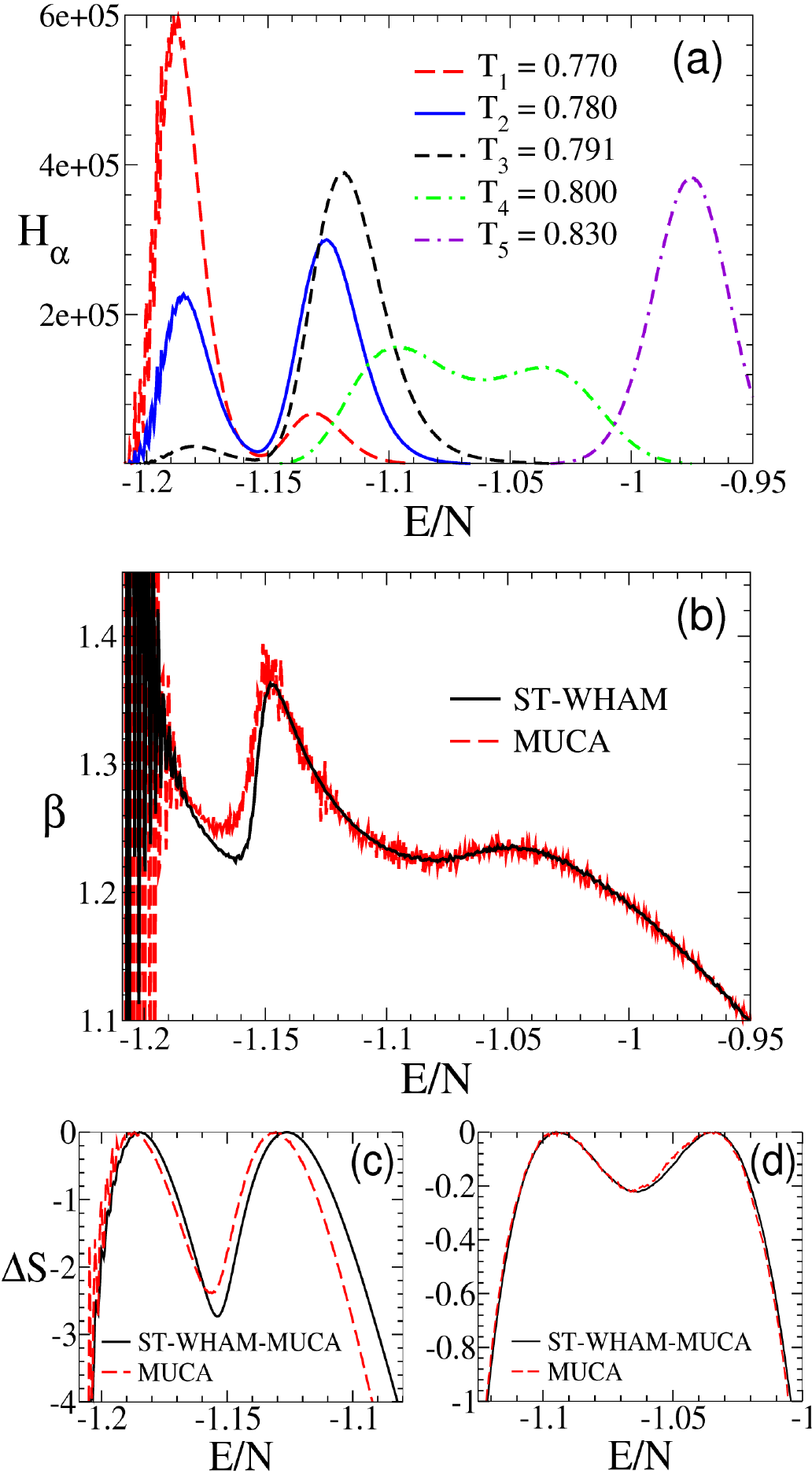}%
\end{center}
\vspace{-0.3cm}
\caption{(a) histograms $H_{\alpha}$; (b) ST-WHAM (solid line) and MUCA (dashed line)
estimates for the inverse temperature; (c) and (d) shifted microcanonical entropies estimated from
multicanonical-like solution of Eq. (\ref{eq:ST}) (solid line) and MUCA
(dashed line) as a function of $E/N$ for lattice size $L=48$.}
\end{figure}


\end{document}